\documentclass[letterpaper, 10 pt, conference]{ieeeconf}
\IEEEoverridecommandlockouts
\usepackage{amsmath,amssymb,amsfonts}
\usepackage{hyperref}
\hypersetup{colorlinks,allcolors=black}
\usepackage{cleveref}
\usepackage{times}

\usepackage{graphicx} %

\usepackage{svg}
\usepackage[font=small,labelfont=bf]{caption}
\usepackage{graphicx}
\usepackage{xcolor}
\usepackage{amssymb}
\usepackage{mathtools}
\usepackage{todonotes}
\let\labelindent\relax
\usepackage{enumitem}
\usepackage{comment}
\usepackage[per-mode=symbol]{siunitx}
\DeclareSIUnit\minute{min}
\usepackage{tikzpagenodes}
\usepackage{algorithm}
\usepackage[noend]{algpseudocode}

\usepackage[style=ieee,natbib=true,backend=biber]{biblatex}
\usepackage{multicol}
\usepackage{booktabs}

\makeatletter
\let\NAT@parse\undefined
\makeatother

\begin{document}

\title{Informative Input Design for Dynamic Mode Decomposition}

\author{
    Joshua Ott\textsuperscript{\rm 1}\thanks{\textsuperscript{\rm 1}Department of Aeronautics \& Astronautics, Stanford University.},
    Mykel J. Kochenderfer\textsuperscript{\rm 1}\thanks{\textsuperscript{\rm 2}Department of Electrical Engineering, Stanford University},
    and Stephen Boyd\textsuperscript{\rm 2}
}

\maketitle

\begin{abstract}
Efficiently estimating system dynamics from data is essential for minimizing data collection costs and improving model performance. This work addresses the challenge of designing future control inputs to maximize information gain, thereby improving the efficiency of the system identification process. We propose an approach that integrates informative input design into the Dynamic Mode Decomposition with control (DMDc) framework, which is well-suited for high-dimensional systems. By formulating an approximate convex optimization problem that minimizes the trace of the estimation error covariance matrix, we are able to efficiently reduce uncertainty in the model parameters while respecting constraints on the system states and control inputs. This method outperforms traditional techniques like Pseudo-Random Binary Sequences (PRBS) and orthogonal multisines, which do not adapt to the current system model and often gather redundant information. %
We validate our approach using aircraft and fluid dynamics simulations to demonstrate the practical applicability and effectiveness of our method. Our results show that strategically planning control inputs based on the current model enhances the accuracy of system identification while requiring less data. Furthermore, we provide our implementation and simulation interfaces as an open-source software package, facilitating further research development and use by industry practitioners.
\end{abstract}

\section{Introduction}\label{sec:introduction}
Estimating system dynamics from data is a fundamental problem in control theory and systems engineering \cite{ljung1998system, tangirala2018principles, keesman2011system, kostelich1992problems, vandenberghe2012convex}. Rapidly learning from limited measurements in high-dimensional systems can significantly reduce the cost associated with expensive real-world data collection while yielding more accurate models. Optimizing the actuation input sequence to gather informative measurements while respecting constraints on the state and control inputs remains an open challenge \cite{kaiser2018sparse}. From the perspective of experimental design, designing future input signals to aid in the identification of the dynamical system can be framed as an information maximization problem \cite{uy2009optimization, wahlberg2010optimal}. This involves perturbing the system in directions that provide high-value information, thereby enhancing the efficiency of the learning process.%

System identification is relevant in almost every science and engineering discipline ranging from aircraft dynamics, to geothermal processes, to financial markets \cite{tangirala2018principles}. When the state space of these real-world environments is very large, reduced order models can be used to capture the most prominent dynamics of the system of interest. One common approach to constructing reduced order models is through Dynamic Mode Decomposition (DMD), a data-driven technique that decomposes complex systems into a set of dynamic modes, capturing essential spatiotemporal patterns \cite{schmid2010dynamic, proctor2016dynamic}. Efficiently reducing uncertainty in the system dynamics model results in less data required to achieve similar predictive model performance while also reducing the costs related to expensive real-world data collection. 

Designing future control inputs to efficiently learn system dynamics is challenging because future state estimates depend on the current model \cite{rojas2007robust, bombois2006least, suzuki2007input, mu2018input}. Minimizing uncertainty in the model derived from data requires understanding how control inputs will affect future states, which in turn requires a reliable model \cite{morelli2016aircraft}. This interdependence might seem circular, but sequentially optimizing inputs significantly reduces model uncertainty more rapidly and with less data than using random input sequences while also satisfying constraints on the resulting states and control inputs \cite{barenthin2008complexity}. While collecting data with random inputs can indeed improve the identification of the underlying model, strategically planning future control inputs based on the current system model often yields larger improvements with less data, enhancing the efficiency of the learning process while also remaining cognizant of constraints on the system states and control inputs \cite{brighenti2009input, fujimoto2018informative}. 

Popular techniques for designing informative inputs such as Pseudo-Random Binary Sequences (PRBS) and orthogonal multisines \cite{rivera2009constrained, morelli2021optimal} do not use the current understanding of the system model. As a result, they often waste time collecting redundant information. Ideally, we would use our current model to identify regions in the parameter space where variance is high. By tailoring future control inputs to our current model, we can more efficiently reduce uncertainty. %

Our proposed approach integrates informative input design into the Dynamic Mode Decomposition with control (DMDc) framework, significantly enhancing the accuracy and efficiency of system identification. By formulating an optimization problem that minimizes the trace of the model covariance matrix, we systematically reduce the uncertainty in our estimates of the model parameters while respecting constraints on the state and control variables. We directly compare our approach to the optimal input design method proposed by \citeauthor{morelli2021optimal} \cite{morelli2021optimal}.

The main focus of our work is to design informative inputs while respecting state and input constraints using our current estimate of the model. For very large state spaces, we can reduce the state space size first through DMDc and then everything else works the same as before.

\begin{figure*}[t]
\centering
    {\includegraphics[width=1\textwidth]{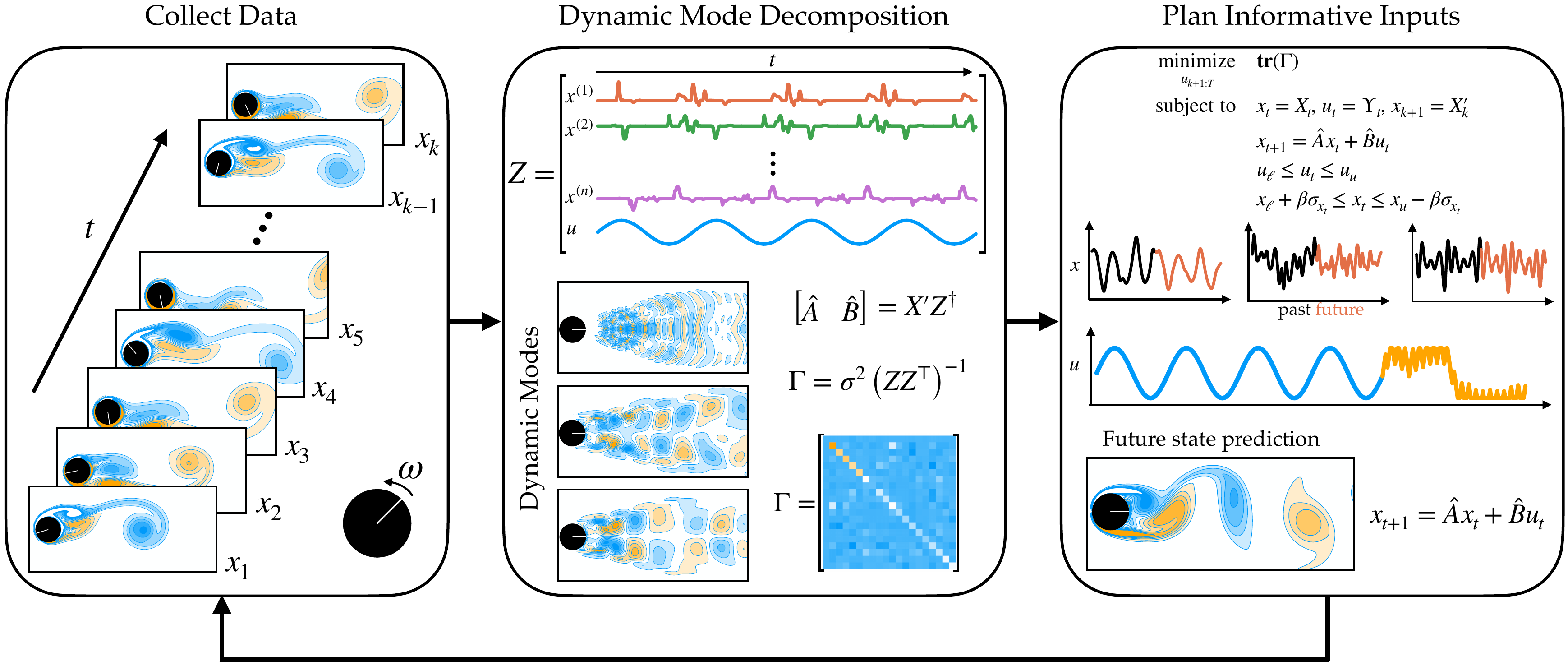}}    
  \caption{Schematic of the informative input design for DMDc pipeline, demonstrated by controlling the rotation of a cylinder in a fluid. Data are collected from the system for the $n$ state variables and $m$ control inputs. Next, DMDc is performed to construct a reduced order model of the system. The covariance matrix $\Gamma$ is constructed and is then used to plan future informative control inputs.} \label{fig:main}
\end{figure*}

The key contributions of this work include introducing a convex approximation of the optimal input design problem, extending our method to handle high-dimensional state spaces through reduced-order models using the DMDc framework, validating our approach with simulations using the WaterLily, Aerobench, and X-Plane aircraft dynamic simulators \cite{weymouth_waterlily, heidlauf2018verification}, and releasing our implementation and simulation interfaces as an open-source software package.\footnote{Code is available at https://github.com/josh0tt/InputOptimization}

\section{Related Work}\label{sec:related_work}
Dynamic Mode Decomposition (DMD) was originally introduced for the modeling of fluid flows \cite{schmid2010dynamic} and was later extended to incorporate the effects of control \cite{proctor2016dynamic}. More generally, DMD can simply be viewed as applying regression techniques for system identification. The model that is produced is only as good as the data that was used to create it. This naturally motivates the problem of optimizing future control inputs to maximally reduce the uncertainty in the identified model. 

The concept of optimal input design to estimate the parameters of a dynamical system is not new and was studied as early as the 1950s \cite{slepian1954estimation, wallis1968optimal, nahi1971design, aoki1970input, gupta1975input}. Recently, there has been a renewed interest in optimal input design \cite{barker2004optimal, hjalmarsson2007optimal, rojas2007robust, jauberthie2006optimal, hjalmarsson2008closed, manchester2010input, gerencser2016adaptive} stemming from advances in sparse system identification methods from large data sets \cite{brunton2016discovering, kaiser2018sparse} as well as in nonlinear optimization and learning-based methods \cite{vincent2010input, noel2017nonlinear, tsiamis2022learning, lee2024active}. 

\citeauthor{goodwin1971optimal} proposed an iterative design algorithm based on gradient descent which seeks to optimize the trace of the covariance matrix with an additional penalty term for violating constraints \cite{goodwin1971optimal}. \citeauthor{barenthin2006input} approached the problem from the frequency domain and used an approximation of the Fisher information matrix as the optimization objective \cite{barenthin2006input}. \citeauthor{babar2016mpc} used model predictive control (MPC) techniques for optimal input design in nonlinear system identification with an Extended Kalman Filter (EKF) to update parameter estimates \cite{babar2016mpc}. 

\citeauthor{subedi2023online} formulated the problem as a nonlinear optimization problem \cite{subedi2023online} where the solution was warm started with the result from a dynamic programming solution developed by \citeauthor{Morelli1993} \cite{Morelli1993}. \citeauthor{wang2020optimal} also formulated the problem as a nonlinear optimization problem and used ant colony optimization to produce a pseudo-random sequence \cite{wang2020optimal}. A particularly interesting recent approach proposed by \citeauthor{morelli2021optimal} uses a nonlinear orthogonal multisine input design technique, generating multiple mutually orthogonal sinusoidal inputs with optimized phase shifts %
\cite{morelli2021optimal}. %
Our experiments compare our method against the method proposed by \citeauthor{morelli2021optimal}.

\section{Problem Formulation}\label{sec:problem_formulation}
Given a linear dynamical system represented by \begin{equation} 
x_{t+1} = A x_t + B u_t + v_t 
\end{equation} where the state at time $t$ is $x_t \in \mathbf{R}^n$, the control input is $u_t \in \mathbf{R}^m$, $A \in \mathbf{R}^{n \times n}$, $B \in \mathbf{R}^{n \times m}$, and $v_t \sim \mathcal{N}(0, \sigma^2\mathrm{I})$ with state constraints $x_{\ell} \leq x_t \leq x_u$ and control constraints $u_{\ell} \leq u_t \leq u_u$. Let \begin{equation}
    X = \begin{bmatrix} \vert & \vert & & \vert \\ x_1 & x_2 & \cdots & x_k \\ \vert & \vert & & \vert \end{bmatrix} \quad X' = \begin{bmatrix} \vert & \vert & & \vert \\ x_2 & x_3 & \cdots & x_{k+1} \\ \vert & \vert & & \vert \end{bmatrix} \nonumber
\end{equation} \begin{equation}
    \Upsilon = \begin{bmatrix} \vert & \vert & & \vert \\ u_1 & u_2 & \cdots & u_k \\ \vert & \vert & & \vert \end{bmatrix} \quad \Theta = \begin{bmatrix} A & B \end{bmatrix} \quad Z = \begin{bmatrix} X \\ \Upsilon \end{bmatrix}
\end{equation} where we collected state and control input data up to time step $k+1$ and $k$ respectively. The linear system model is \begin{equation} 
X' = AX + B\Upsilon + V = \begin{bmatrix} A & B \end{bmatrix} \begin{bmatrix} X \\ \Upsilon \end{bmatrix} + V = \Theta Z + V
\end{equation} where $V \in \mathbf{R}^{n \times k}$ is the stacked noise matrix. We can construct estimates of $A$ and $B$, which we denote $\hat{A}$ and $\hat{B}$, by solving the regression problem with the Frobenius norm: \begin{equation}
    \underset{\hat{\Theta}}{\text{minimize}} \left\lVert \hat{\Theta} Z - X' \right\rVert_F^2
\end{equation} where $\hat{\Theta} = \begin{bmatrix} \hat{A} & \hat{B} \end{bmatrix}$. This optimization problem can be solved using the least squares method, yielding: \begin{equation}
    \hat{\Theta} = X' Z^\dagger \label{eq:theta_pinv}
\end{equation} where $Z^\dagger$ is the pseudo-inverse of $Z$. If $\Gamma = \sigma^2 \left(Z Z^{\top}\right)^{-1}$, the covariance of $\hat{\Theta}$ is given by: \begin{equation}
   \mathrm{Cov}(\hat{\Theta}) = \mathrm{I}_n \otimes \Gamma
\end{equation} %
as shown in \cref{sec:appendix} and is equivalently the covariance of the least squares parameter estimate \cite{hastie2009elements}. The trace of $\mathrm{Cov}(\hat{\Theta})$ is a measure of the total variance of the estimates of $ \hat{A} $ and $ \hat{B} $. Notice that minimizing the trace of $\mathrm{Cov}(\hat{\Theta})$ is equivalent to minimizing the trace of $ \Gamma $ as shown in \cref{sec:appendix}. The root-mean-square-error (RMSE) between the estimated $\hat{\Theta}$ and the true $\Theta$ is given by \begin{equation}
     \mathrm{RMSE}  =  \sqrt{\frac{1}{n(n+m)} \mathbf{tr}(\mathrm{Cov}(\hat{\Theta}))} \label{eq:RMSE}
\end{equation}
where $\mathbf{tr}(\mathrm{Cov}(\hat{\Theta}))$ represents the trace of the covariance matrix. Since we have assumed that the process noise $v_t$ is zero mean Gaussian noise, we can compute the variance of forward propagating the dynamics into the future as  \begin{equation}
    \Sigma_{x_{k+1}} = A \Sigma_{x_k} A^\top + \sigma^2\mathrm{I}.
\end{equation}
This recursive equation allows us to estimate how uncertainty in the state evolves over time due to both the dynamics of the system and the influence of the process noise. We have $\sigma_{x_t}^2 = \text{diag}(\Sigma_{x_t})$.%

To minimize our uncertainty in these estimates, we pose the following optimization problem. Let $t=1,\dots,k+1$ denote the time steps for which we have collected data to construct estimates of $\hat{A}$ and $\hat{B}$. We then want to plan future control inputs $ u_t $ from $t=k+1,\dots,T$: \small \begin{align} 
    \underset{u_{k+1:T}}{\text{minimize}} \quad & \mathbf{tr}(\Gamma) \label{eq:control_problem} \\
    \text{subject to} \quad & x_t = X_t, \hspace{0.15em} u_t = \Upsilon_t, \hspace{0.15em} x_{k+1} = X'_k && t = 1,\dots, k \nonumber \\
    & x_{t+1} = \hat{A} x_t + \hat{B} u_t  && t = k+1,\dots, T \nonumber \\
    & u_{\ell} \leq u_t \leq u_u && t = k+1,\dots, T \nonumber \\
    & x_{\ell} + \beta \sigma_{x_t} \leq x_t \leq x_u - \beta \sigma_{x_t} && t = k+2,\dots, T \nonumber
\end{align} \normalsize where the variables are $u$ and $x$. The lower and upper constraints on the state and control inputs are $u_{\ell}, x_{\ell}, u_u,$ and $x_u$ respectively. %
The standard deviation is scaled by $\beta$ for the upper and lower bound state constraints. Everything in \cref{eq:control_problem} is convex, except for the objective $\mathbf{tr}(\Gamma)$. 

The solution to this problem will guide the selection of control inputs that not only steer the system state in a desired manner but also do so in a way that maximally reduces the uncertainty in our estimates of system dynamics resulting in more efficient system identification. %

\subsection{Dynamic Mode Decomposition with Control}
\citeauthor{proctor2016dynamic} presented Dynamic Mode Decomposition with control (DMDc) to produce a reduced order model motivated by the fact that when $n \gg 1$, computing the pseudo-inverse in \cref{eq:theta_pinv} can be computationally prohibitive \cite{proctor2016dynamic}. 

DMDc first reduces the dimension of the input space through the singular value decomposition (SVD). The SVD of $Z$ results in \begin{equation}
    Z = U\Sigma V^\ast \approx \tilde{U} \tilde{\Sigma}^{-1} \tilde{V}^{\ast}
\end{equation} where the right side is a low rank approximation using the first $p$ singular values %
\cite{eckart1936approximation}. We have $\bar{\Theta} = X'\tilde{V}\tilde{\Sigma}^{-1}\tilde{U}^\ast$ with $\tilde{U} = \begin{bmatrix}
    \tilde{U}_1^\ast & \tilde{U}_2^\ast
\end{bmatrix}^{\top}$ giving \begin{equation}
   \begin{bmatrix} \bar{A} & \bar{B} \end{bmatrix} = \begin{bmatrix} X'\tilde{V}\tilde{\Sigma}^{-1}\tilde{U}_1^\ast & X'\tilde{V}\tilde{\Sigma}^{-1}\tilde{U}_2^\ast \end{bmatrix}.
\end{equation} 

We then apply the SVD to the output space $X' = \hat{U}\hat{\Sigma}\hat{V}^\ast$ with truncation value $r$ and $\hat{U} \in \mathbf{R}^{n \times r}$, $\hat{\Sigma} \in \mathbf{R}^{r \times r}$, $\hat{V}^\ast \in \mathbf{R}^{r \times k}$. The dimension of $x$ can be reduced through the transformation $x = \hat{U}\tilde{x}$ and the reduced order dynamics are defined by \begin{equation}
    \tilde{A} = \hat{U}^\ast X' \tilde{V} \tilde{\Sigma}^{-1} \tilde{U}_1^\ast \hat{U} \label{eq:tilde_A}
\end{equation} \begin{equation}
    \tilde{B} = \hat{U}^\ast X' \tilde{V} \tilde{\Sigma}^{-1} \tilde{U}_2^\ast \label{eq:tilde_B}
\end{equation} with $\tilde{A} \in \mathbf{R}^{r \times r}$ and $\tilde{B} \in \mathbf{R}^{r \times m}$ giving $\hat{\Theta} = \begin{bmatrix}
    \tilde{A} & \tilde{B}
\end{bmatrix}$ when $n \gg 1$. We have reduced the dimension of the problem from $n$ to $r$. The reduced order dynamics are: \begin{equation}
    \tilde{x}_{k+1} = \tilde{A}\tilde{x}_k + \tilde{B}u_k
\end{equation} with $\tilde{Z} = \begin{bmatrix}
    \tilde{X} \\ \Upsilon
\end{bmatrix}$ where $\tilde{X} = \hat{U}^\ast X$. The problem in \cref{eq:control_problem} is then \small \begin{align} 
    \underset{u_{k+1:T}}{\text{minimize}} \quad & \mathbf{tr}(\tilde{\Gamma}) \label{eq:reduced_control_problem} \\
    \text{subject to} \quad & \tilde{x}_t = \tilde{X}_t, \hspace{0.15em} u_t = \Upsilon_t, \hspace{0.15em} \tilde{x}_{k+1} = \tilde{X}'_k && t = 1,\dots, k \nonumber \\
    & \tilde{x}_{t+1} = \tilde{A} \tilde{x}_t + \tilde{B} u_t && t = k+1,\dots, T \nonumber \\
    & u_{\ell} \leq u_t \leq u_u && t = k+1,\dots, T \nonumber \\
    & \tilde{x}_{\ell} + \beta \sigma_{\tilde{x}_t} \leq \tilde{x}_t \leq \tilde{x}_u - \beta \sigma_{\tilde{x}_t} && t = k+2,\dots, T \nonumber
\end{align} \normalsize where $\tilde{\Gamma} = \sigma^2 \left(\tilde{Z}\tilde{Z}^\top \right)^{-1}$.

\section{Convex Approximation}\label{sec:methods}
\begin{figure*}[t]
\centering
    {\includegraphics[width=1\textwidth]{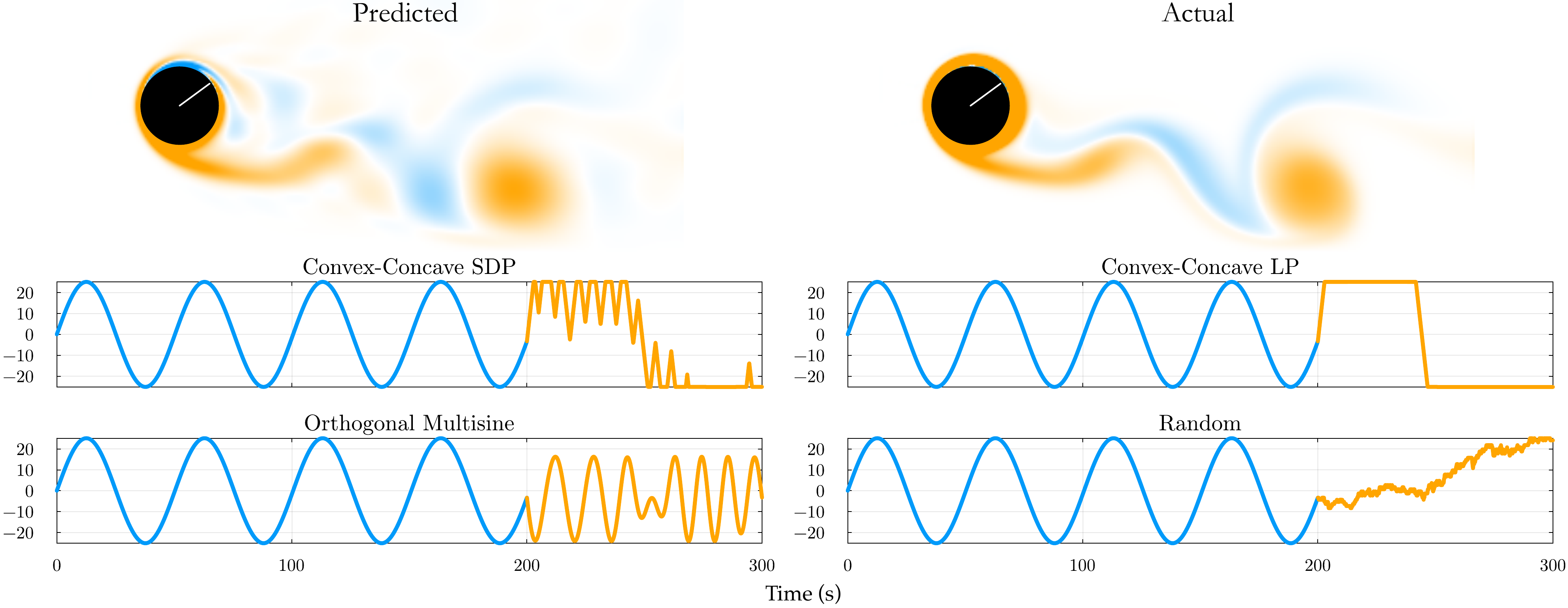}}  
  \caption{Predicted and actual vorticity field for fluid flow past a cylinder at $\mathrm{Re}=100$ after collecting data from the informative inputs of the SDP method. The four plots show the control inputs planned by each of the four methods. Blue indicates the initial data set and orange indicates the informative inputs that were designed.} \label{fig:cylinder}
\end{figure*}

As previously mentioned, \cref{eq:control_problem} and \cref{eq:reduced_control_problem} are convex, except for the objective. The convex-concave procedure (CCP) is a useful heuristic for finding a local optimum by iteratively solving convex optimization problems \cite{lipp2016variations, boyd2016mimo, lipp2016antagonistic}. The CCP replaces concave terms with a convex upper bound, and then solves the resulting convex problem.

We can linearize the term $W = \Gamma^{-1} = \sigma^{-2} Z Z^{\top}$ around the current matrix $Z_c$ which gives: \begin{equation}
    \sigma^{2} \hat W = Z_c Z_c^{\top} +  Z_c(Z-Z_c)^{\top} + (Z-Z_c) Z_c^{\top}.
    \label{eq:w_hat}
\end{equation} This is affine in $Z$, hence the variables $u,x$. We can then minimize $\mathbf{tr}(\hat{W}^{-1})$ instead of $\mathbf{tr}(W^{-1})$ in each iteration of the CCP. The problem then becomes: \begin{align} 
    \underset{u_{k+1:T}}{\text{minimize}} \quad & \mathbf{tr}(\hat{W}^{-1}) \label{eq:sdp} \\
    \text{subject to} \quad & x_{k+2:T} \in \mathcal{X}, \hspace{0.25em} u_{k+1:T} \in \mathcal{U} \nonumber 
\end{align} where $\mathcal{X}$ and $\mathcal{U}$ represent the set of all feasible states and controls given by the constraints in \cref{eq:control_problem}. 

The mapping $Z \to W = \sigma^{-2}ZZ^{\top}$ is matrix convex, making $W(Z) \succeq \hat W(Z)$ for all $Z$. This implies that $W(Z)^{-1} \preceq (\hat W(Z)) ^{-1}$  for all $Z$ and therefore the linearized objective $\mathbf{tr}(\hat{W}^{-1})$ is everywhere above the true objective $\mathbf{tr}(W^{-1})$. 

As a result, we can simply iterate solving the convex problem with objective $\mathbf{tr}(\hat{W}^{-1})$ without the need to line search, update trust regions, or tune hyper parameters.

\subsection{Linear Program}
Alternatively, we can minimize $- \mathbf{tr}(\hat{W})$, which converts the problem from a semidefinite program to a linear program, allowing for an even more efficient solution time. The $- \mathbf{tr}(\hat{W})$ objective is not equivalent to minimizing $\mathbf{tr}(\hat{W}^{-1})$; however, it has been shown to be a useful surrogate objective on large problems \cite{ott2024approximate}. 

It is important to note that minimizing \(\mathbf{tr}(W^{-1})\) penalizes small eigenvalues of \(W\), ensuring that estimation errors are balanced across all directions. In contrast, the \(-\mathbf{tr}(W)\) objective can result in a few large eigenvalues dominating the objective, while others remain small. As a result, this could concentrate information gain in a single direction, resulting in poor performance in orthogonal directions. However, our empirical results demonstrate that this is not an issue in practice as long as the small initial data set used to construct estimates of $\hat{A}$ and $\hat{B}$ have reasonable excitation across each of the control inputs. 

\begin{algorithm}[t]
\caption{Informative Input Design}
\label{alg:informative_input_design}
\begin{algorithmic}[1]
\While{true}
    \If{$n \gg 1$}
        \State $\hat{\Theta} = \begin{bmatrix} \tilde{A} & \tilde{B} \end{bmatrix}$ \Comment{Equations \cref{eq:tilde_A}, \cref{eq:tilde_B}}
    \Else
        \State $\hat{\Theta} = X' Z^\dagger$  \Comment{Equation \cref{eq:theta_pinv}}
    \EndIf

    \State $\hat W \leftarrow \mathrm{linearize}(Z)$ \Comment{Equation \cref{eq:w_hat}}
    
    \State $u_{k+1:T} = \mathrm{minimize}(\hat W)$ \Comment{Equation \cref{eq:sdp}}
    \State Apply the optimized control inputs $u_{k+1:T}$
    \State Collect new data $X_{\mathrm{new}}, X'_{\mathrm{new}}, \Upsilon_{\mathrm{new}}$ and update %
    \State $\Delta t = T - k$
    \State $k, \; T = T, \; T+\Delta t$
\EndWhile
\end{algorithmic}
\end{algorithm}

\subsection{Iterative Input Design}
The iterative nature of the input design described in \cref{alg:informative_input_design} is fundamental to the effectiveness of the procedure. Our method emphasizes continuous improvement through repeated system identification epochs. As illustrated in \cref{fig:main}, we can sequentially solve \cref{eq:control_problem} and \cref{eq:reduced_control_problem} as new data is collected online. %
At each iteration, we use the current estimates of $\hat{A}$ and $\hat{B}$ to generate a new control input trajectory. These inputs are then applied to the system, and the resulting data is used to update our estimates of $\hat{A}$ and $\hat{B}$. This process involves concatenating the most recent measurements with previous data, thereby refining our model iteratively. This iterative input design strategy ensures that we account for the most recent measurements to target areas of the parameter space with the greatest variance.

\section{Results}\label{sec:results}
\begin{figure*}[t]
\centering
    {\includegraphics[width=1\textwidth]{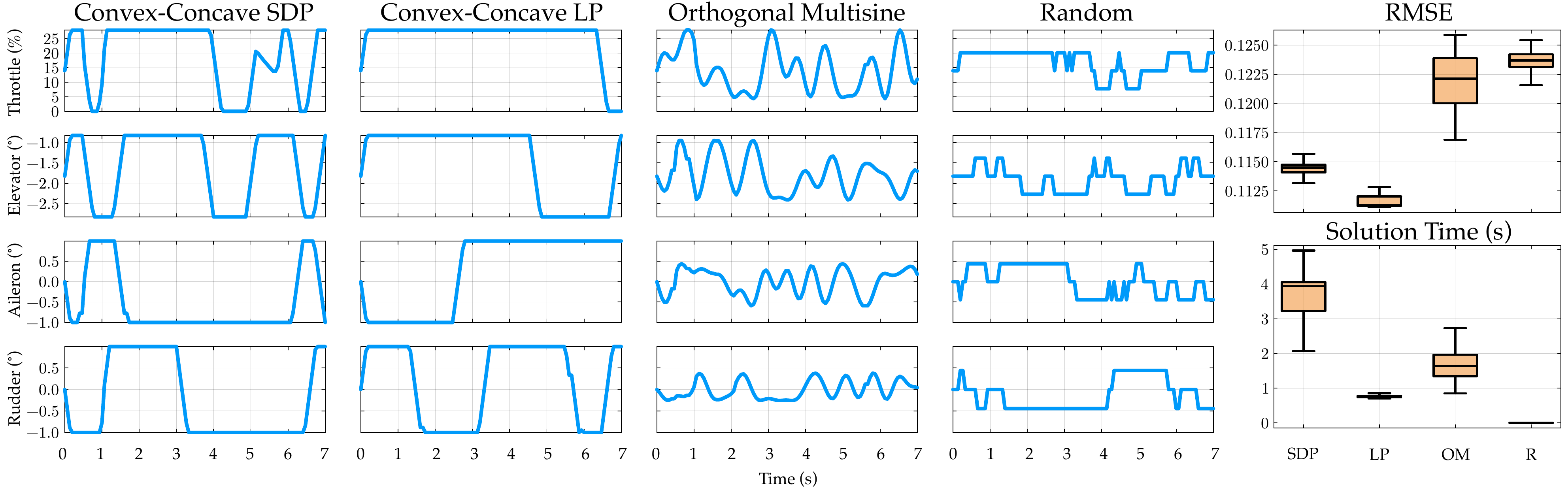}}    
  \caption{Example of the control inputs generated by each of the four methods for the F-16 Aerobench simulations. The right plots show the objective and solution time comparison of the four methods across 100 simulation runs.} \label{fig:combined_plot}
\end{figure*}

We experimentally validated our method in three simulation environments and benchmarked against two common informative input design techniques. The first is the orthogonal multisine method, which generates inputs that are mutually orthogonal in both the time domain and the frequency domain by solving a nonlinear optimization problem \cite{morelli2021optimal}. The second is a random input signal. Our methods are the Convex-Concave semidefinite program (SDP) that uses the linearized objective $\mathbf{tr}(\hat{W}^{-1})$ and the linear program (LP) that uses the $- \mathbf{tr}(\hat{W})$ objective. 

To achieve fair comparison between the three methods, we enforce the same amplitude and time constraints \cite{morelli2016aircraft}. We determine the maximum change in control input $\delta_{\mathrm{max}}$ constraint based on the largest frequency used in the orthogonal multisine method. We then apply this $\delta_{\mathrm{max}}$ constraint to our method as well as the random input method. The random input method changes the input by $\delta_{\mathrm{max}}$ at each time step and is limited to the same amplitude constraint. We compared this random method against a classical filtered pseudo-random binary sequence (PRBS) and found that it performed better so we selected it as the second baseline. 

\subsection{Fluid Flow Past a Cylinder}
We use the WaterLily.jl fluid simulator to control the rotation of a cylinder in horizontal flow at a Reynold's number of $100$ \cite{weymouth_waterlily}. The initial dataset collected at $10$ Hz uses a sinusoidal rotation pattern of the cylinder for $50$ seconds. The size of the vorticity field state space is $n = 51{,}200$. The four methods then plan the future rotation angles of the cylinder for the next $50$ seconds. 

\Cref{fig:cylinder} shows a comparison of the control inputs generated from each of the methods. We can see that the Convex-Concave SDP solution tends to have more jagged oscillations, while the LP is a bit flatter. The random method behaves as expected with consistent oscillations limited by $\delta_{\mathrm{max}}$ and the orthogonal multisine method provides smooth peaks at varying frequencies. 

We also show the predicted state of the vorticity field $10$ seconds into the future after collecting data with informative inputs designed by the SDP method.

\subsection{F-16 Dynamics}
To demonstrate the capability of designing informative inputs for multi-input multi-output systems, we use the Aerobench simulator to simulate the nonlinear dynamics of an F-16 based on the model by \citeauthor{morelli1998global} \cite{morelli1998global}. The state space consists of the velocity, angle of attack, angle of sideslip, roll, pitch, yaw, roll rate, pitch rate, yaw rate, east position, north position, altitude, and engine power setting. The controls are the throttle percentage, elevator, aileron and rudder deflections. 

The initial data consisted of $250$ seconds of simulated flight data collected at $15$ Hz. \Cref{fig:combined_plot} shows the control inputs created by each of the four methods. We also compare the objective value and runtime comparison over 100 simulation runs for each of the methods. The objective values were computed by taking the control inputs planned by each of the methods and then executing them in the Aerobench simulator to collect the true output data. %
The RMSE of $\hat{\Theta}$ was computed with \cref{eq:RMSE} after collecting the data with each of the informative input design methods. %

We see that the SDP and LP methods outperform the other methods by consistently achieving lower RMSE values while performing relatively similarly to each other with notably less variance in the LP performance. We see that the LP method also achieves better runtime performance compared to the orthogonal multisine method. The random method performs the worst in terms of objective value, but has the lowest runtime. The key takeaway here is that the $- \mathbf{tr}(\hat{W})$ objective appears to be a suitable surrogate for the $\mathbf{tr}(\hat{W}^{-1})$ by providing similar performance while decreasing runtime by over $50\%$. Additionally, we see the importance of considering the current model when planning future control inputs in order to target regions of the state and control space where the model has larger variance.

\begin{figure*}[t]
\centering
    {\includegraphics[width=1\textwidth]{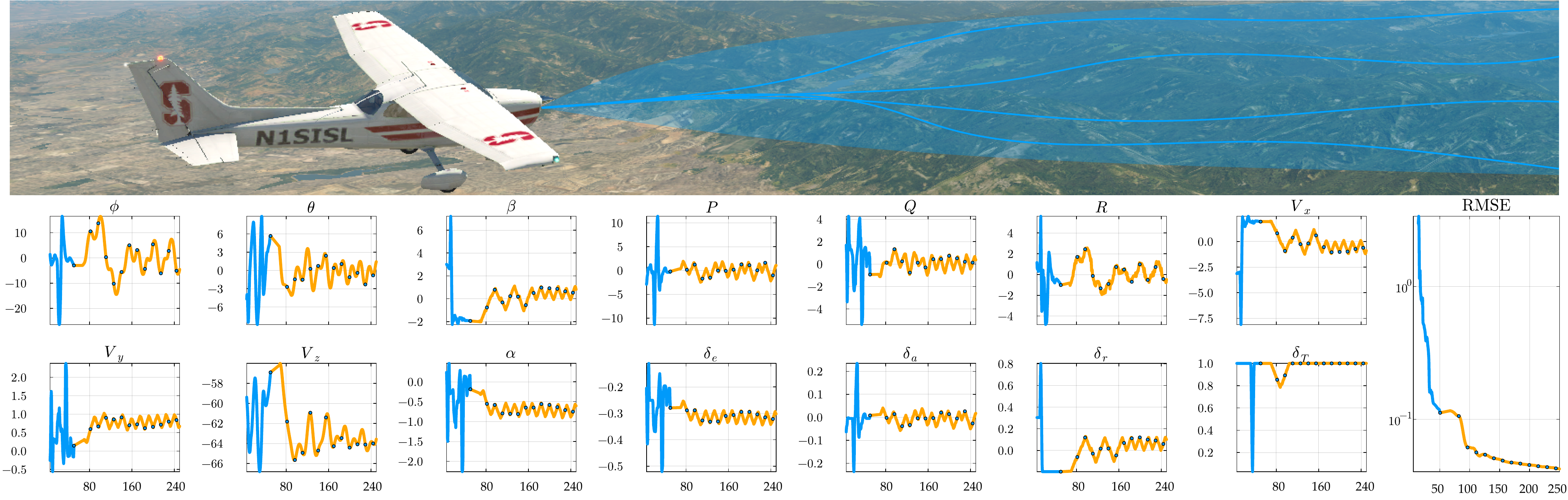}}    
  \caption{Top: image of the Cessna 172 aircraft used for the experiment in the X-Plane simulator with illustrations showing possible future trajectories of the system with uncertainty bounds. Bottom: plots of the state variables throughout the experiment (roll, pitch, sideslip, roll rate, pitch rate, yaw rate, velocity in the x,y,z directions, angle of attack, elevator, aileron, rudder, and throttle positions). Blue: initial data. Orange: data collected during execution of informative inputs. Markers: indicate when each system identification replanning epoch occurred.} \label{fig:xplane}
\end{figure*}

\subsection{Real-Time Example in X-Plane}
Many real-world applications involve situations where the model needs to be updated in real time and new control inputs planned for and executed in an online fashion. To demonstrate the usefulness of our method in real-world situations, we have integrated our method with the X-Plane simulator. X-Plane is a highly realistic flight simulator widely used for both pilot training and research purposes. By interfacing our control input design methods with X-Plane, we were able to test the effectiveness of our approach in a dynamic environment. This integration allows for real-time updates and adjustments to control inputs based on the current state of the aircraft, showcasing the practical applicability of our method in scenarios that require adaptive and responsive control strategies. This example demonstrates the speed of our method to update the model estimate and replan while running onboard the aircraft.

The initial data was collected by having a human fly the airplane for $50$ seconds with data collected at $2$ Hz. The LP method was then run to plan control inputs $25$ seconds into the future. While executing the control inputs, the future state predictions were updated based on the current state of the aircraft. Once the control inputs were executed, the dynamics model was updated with the newly collected data and then a new sequence of control inputs were planned. 

\Cref{fig:xplane} shows the results from the real-time demonstration using a Cessna 172 aircraft. There was no knowledge of the system given to the LP method other than the initial $50$ seconds of data collected.

For this real-time demonstration, we added additional constraints so that the system would try to end each $25$ second maneuver back at the initial trim condition (zero roll, pitch, yaw, sideslip, etc.). This is consistent with the approach used by \citeauthor{morelli2021optimal} to avoid the aircraft from deviating too far from the flight region of interest. The average replanning time for the LP method during these experiments was 0.20 seconds. All of the results presented in this section were run on a computer with an M1 Max and 64 GB of RAM.

\section{Discussion}\label{sec:conclusion}
We demonstrated a powerful technique to efficiently design informative control inputs for dynamic system identification, with a particular focus on high dimensional systems. Our approach integrates informative input design into the Dynamic Mode Decomposition with control (DMDc) framework by formulating the problem as an approximate convex optimization problem. This formulation allows us to systematically reduce model uncertainty and improve predictive accuracy while adhering to practical constraints on the resulting state estimates, control input amplitudes, slew rates, trim conditions, as well as the possibility to extend to a variety of other practical constraints.

We validated our methods through extensive experiments in three different simulation environments. In the fluid flow past a cylinder experiment, we demonstrated the ability of our method to handle high dimensional systems. In the Aerobench simulator with F-16 dynamics, we showcased the effectiveness of our approach in handling complex, multi-input multi-output control scenarios, highlighting the advantages of our proposed methods in terms of objective value performance and solution times. Finally, we integrated our method with the X-Plane simulator to demonstrate real-time applicability, emphasizing its potential for adaptive and responsive control in real-world environments.

The results from these simulations underscore the practical utility of our approach in various real-world applications, from aircraft control to fluid dynamics. Our methods not only significantly reduce model uncertainty with less data, but also support real-time applications.

Future work will explore enhancements to our input design framework, including extensions to handle more complex nonlinear systems, and applications in other domains such as robotics, climate modeling, and financial systems. The integration of our methods with real-time control systems holds promise for advancing the state-of-the-art in system identification and control, paving the way for more intelligent, data-driven approaches to managing and understanding complex dynamical systems. As data continues to become more abundant, the question becomes not only what information is contained in your data, but often, and more importantly -- what is not. Our contribution provides a principled approach to maximally leverage the data that we have collected by informing what new data needs to be collected. In doing so, we believe this will allow for the development of better informed data driven models while also maintaining the efficiency required for real-time applications.

\section{Acknowledgments}
We would like to thank Lauren Ho-Tseung for her contributions in developing this method.

\printbibliography

\onecolumn
\section{Appendix}\label{sec:appendix}
\subsection{Derivation of $ \mathrm{Cov}(\hat{\Theta})$}
We have: \[
\mathbf{Z} = 
\begin{bmatrix}
x_1^{(1)} & x_1^{(2)} & \cdots & x_1^{(k)} \\
x_2^{(1)} & x_2^{(2)} & \cdots & x_2^{(k)} \\
\vdots & \vdots & \ddots & \vdots \\
x_n^{(1)} & x_n^{(2)} & \cdots & x_n^{(k)} \\
u_1^{(1)} & u_1^{(2)} & \cdots & u_1^{(k)} \\
\vdots & \vdots & \ddots & \vdots \\
u_m^{(1)} & u_m^{(2)} & \cdots & u_m^{(k)}
\end{bmatrix} \qquad \mathbf{Z}^{\top} = 
\begin{bmatrix}
x_1^{(1)} & x_2^{(1)} & \cdots & x_n^{(1)} & u_1^{(1)} & \cdots & u_m^{(1)} \\
x_1^{(2)} & x_2^{(2)} & \cdots & x_n^{(2)} & u_1^{(2)} & \cdots & u_m^{(2)} \\
\vdots & \vdots & \ddots & \vdots & \vdots & \ddots & \vdots \\
x_1^{(k)} & x_2^{(k)} & \cdots & x_n^{(k)} & u_1^{(k)} & \cdots & u_m^{(k)}
\end{bmatrix}
\]
where \(\mathbf{Z} \in \mathbb{R}^{(n+m) \times k}\) and \(\mathbf{Z}^{\top} \in \mathbb{R}^{k \times (n+m)}\). We then have 

\begin{equation}
    \mathbf{Z}\mathbf{Z}^{\top} =
    \begin{bmatrix}
    \sum_{i=1}^k x_1^{(i)} x_1^{(i)} & \sum_{i=1}^k x_1^{(i)} x_2^{(i)} & \cdots & \sum_{i=1}^k x_1^{(i)} u_m^{(i)} \\
    \sum_{i=1}^k x_2^{(i)} x_1^{(i)} & \sum_{i=1}^k x_2^{(i)} x_2^{(i)} & \cdots & \sum_{i=1}^k x_2^{(i)} u_m^{(i)} \\
    \vdots & \vdots & \ddots & \vdots \\
    \sum_{i=1}^k u_m^{(i)} x_1^{(i)} & \sum_{i=1}^k u_m^{(i)} x_2^{(i)} & \cdots & \sum_{i=1}^k u_m^{(i)} u_m^{(i)}
    \end{bmatrix}
\end{equation} where \(\mathbf{Z}\mathbf{Z}^{\top} \in \mathbb{R}^{(n+m) \times (n+m)}\).

Recall, we have $\hat{\Theta} = \begin{bmatrix} \hat{A} & \hat{B} \end{bmatrix} \in \mathbb{R}^{n \times (n+m)}$ with $\hat{\Theta} = X' Z^\dagger$. We can vectorize $\hat{\Theta}$ to setup a classical regression problem. Let $\hat{\beta} = \text{vec}(\hat{\Theta})$. We then define \(\mathbf{\Phi}\) as 
\begin{equation}
\mathbf{\Phi} = 
\begin{bmatrix}
x_1^{(1)} & x_2^{(1)} & \cdots & u_m^{(1)} & 0 & 0 & \cdots & 0 & 0 & \cdots & 0 & 0 \\
0 & 0 & \cdots & 0 & x_1^{(1)} & x_2^{(1)} & \cdots & u_m^{(1)} & 0 & \cdots & 0 & 0 \\
\vdots & \vdots & \ddots & \vdots & \vdots & \vdots & \ddots & \vdots & \vdots & \ddots & \vdots & \vdots \\
0 & 0 & \cdots & 0 & 0 & 0 & \cdots & 0 & x_1^{(1)} & x_2^{(1)} & \cdots & u_m^{(1)} \\
x_1^{(2)} & x_2^{(2)} & \cdots & u_m^{(2)} & 0 & 0 & \cdots & 0 & 0 & \cdots & 0 & 0 \\
0 & 0 & \cdots & 0 & x_1^{(2)} & x_2^{(2)} & \cdots & u_m^{(2)} & 0 & \cdots & 0 & 0 \\
\vdots & \vdots & \ddots & \vdots & \vdots & \vdots & \ddots & \vdots & \vdots & \ddots & \vdots & \vdots \\
0 & 0 & \cdots & 0 & 0 & 0 & \cdots & 0 & x_1^{(2)} & x_2^{(2)} & \cdots & u_m^{(2)} \\
\vdots & \vdots & \ddots & \vdots & \vdots & \vdots & \ddots & \vdots & \vdots & \ddots & \vdots & \vdots \\
x_1^{(k)} & x_2^{(k)} & \cdots & u_m^{(k)} & 0 & 0 & \cdots & 0 & 0 & \cdots & 0 & 0 \\
0 & 0 & \cdots & 0 & x_1^{(k)} & x_2^{(k)} & \cdots & u_m^{(k)} & 0 & \cdots & 0 & 0 \\
\vdots & \vdots & \ddots & \vdots & \vdots & \vdots & \ddots & \vdots & \vdots & \ddots & \vdots & \vdots \\
0 & 0 & \cdots & 0 & 0 & 0 & \cdots & 0 & x_1^{(k)} & x_2^{(k)} & \cdots & u_m^{(k)}
\end{bmatrix} 
\end{equation}
\begin{equation}
    \mathbf{\Phi}^{\top} = 
\begin{bmatrix}
x_1^{(1)} & 0 & \cdots & 0 & x_1^{(2)} & 0 & \cdots & 0 & \cdots & x_1^{(k)} & 0 & \cdots & 0 \\
x_2^{(1)} & 0 & \cdots & 0 & x_2^{(2)} & 0 & \cdots & 0 & \cdots & x_2^{(k)} & 0 & \cdots & 0 \\
\vdots & \vdots & \ddots & \vdots & \vdots & \vdots & \ddots & \vdots & \ddots & \vdots & \vdots & \ddots & \vdots \\
u_m^{(1)} & 0 & \cdots & 0 & u_m^{(2)} & 0 & \cdots & 0 & \cdots & u_m^{(k)} & 0 & \cdots & 0 \\
0 & x_1^{(1)} & \cdots & 0 & 0 & x_1^{(2)} & \cdots & 0 & \cdots & 0 & x_1^{(k)} & \cdots & 0 \\
0 & x_2^{(1)} & \cdots & 0 & 0 & x_2^{(2)} & \cdots & 0 & \cdots & 0 & x_2^{(k)} & \cdots & 0 \\
\vdots & \vdots & \ddots & \vdots & \vdots & \vdots & \ddots & \vdots & \ddots & \vdots & \vdots & \ddots & \vdots \\
0 & u_m^{(1)} & \cdots & 0 & 0 & u_m^{(2)} & \cdots & 0 & \cdots & 0 & u_m^{(k)} & \cdots & 0 \\
\vdots & \vdots & \ddots & \vdots & \vdots & \vdots & \ddots & \vdots & \ddots & \vdots & \vdots & \ddots & \vdots \\
0 & 0 & \cdots & x_1^{(1)} & 0 & 0 & \cdots & x_1^{(2)} & \cdots & 0 & 0 & \cdots & x_1^{(k)} \\
0 & 0 & \cdots & x_2^{(1)} & 0 & 0 & \cdots & x_2^{(2)} & \cdots & 0 & 0 & \cdots & x_2^{(k)} \\
\vdots & \vdots & \ddots & \vdots & \vdots & \vdots & \ddots & \vdots & \ddots & \vdots & \vdots & \ddots & \vdots \\
0 & 0 & \cdots & u_m^{(1)} & 0 & 0 & \cdots & u_m^{(2)} & \cdots & 0 & 0 & \cdots & u_m^{(k)}
\end{bmatrix}
\end{equation} where \(\mathbf{\Phi} \in \mathbb{R}^{kn \times n(n+m)}\) and \(\mathbf{\Phi}^{\top} \in \mathbb{R}^{n(n+m) \times kn}\). 

We have \begin{equation}
    y = \text{vec}(X^{\prime}) = \Phi \hat{\beta}
\end{equation} with the regression covariance matrix given by \begin{equation}
    \text{Cov}(\hat{\beta}) = \text{Cov}(\hat{\Theta}) = \hat{\sigma}^2(\Phi^{\top} \Phi)^{-1}
\end{equation}

From the above definitions we have
\[
\mathbf{\Phi}^{\top}\mathbf{\Phi} = 
\begin{bmatrix}
\mathbf{Z}\mathbf{Z}^{\top} & 0 & \cdots & 0 \\
0 & \mathbf{Z}\mathbf{Z}^{\top} & \cdots & 0 \\
\vdots & \vdots & \ddots & \vdots \\
0 & 0 & \cdots & \mathbf{Z}\mathbf{Z}^{\top}
\end{bmatrix}
\]
This matrix is block diagonal with \(\mathbf{Z}\mathbf{Z}^{\top}\) repeated \(n\) times.

\[
\mathbf{\Phi}^{\top}\mathbf{\Phi} = \mathbf{I}_n \otimes \mathbf{Z}\mathbf{Z}^{\top}
\]

We know \(\mathbf{\Gamma} = \hat{\sigma}^2 (\mathbf{Z}\mathbf{Z}^{\top})^{-1}\) and

\[
\text{Cov}(\hat{\mathbf{\Theta}}) = \hat{\sigma}^2 (\mathbf{\Phi}^{\top}\mathbf{\Phi})^{-1} = \hat{\sigma}^2 (\mathbf{I}_n \otimes \mathbf{Z}\mathbf{Z}^{\top})^{-1}
\] Using the property of the Kronecker product:

\[
(A \otimes B)^{-1} = A^{-1} \otimes B^{-1}
\]

\[
\text{Cov}(\hat{\mathbf{\Theta}}) = \hat{\sigma}^2 (\mathbf{I}_n \otimes (\mathbf{Z}\mathbf{Z}^{\top})^{-1})
\]

\[
= \mathbf{I}_n \otimes \hat{\sigma}^2 (\mathbf{Z}\mathbf{Z}^{\top})^{-1}
\]

\[
= \mathbf{I}_n \otimes \mathbf{\Gamma}
\]

\[
\text{Cov}(\hat{\mathbf{\Theta}}) = \mathbf{I}_n \otimes \mathbf{\Gamma}
\]

\subsection{RMSE Derivation}
The RMSE is defined as:

\[
\mathrm{RMSE} = \sqrt{\frac{1}{n (n + m)} \sum_{i=1}^{n} \sum_{j=1}^{n+m} (\hat{\Theta}_{i,j} - \Theta_{i,j})^2}
\]

We have
   \[
   \Gamma = \sigma^2 (Z Z^{\top})^{-1}
   \]
   
   \[\mathrm{Cov}(\hat{\Theta}) = \mathrm{I}_n \otimes \Gamma   \]

   \[
   \mathrm{Var}(\hat{\Theta}_{i,j}) = \left[\mathrm{I}_n \otimes \Gamma\right]_{i,j}
   \]

If \(\hat{\Theta}\) is an unbiased estimator, the expected value of the squared error is the variance:
   \[
   \mathbb{E}[(\hat{\Theta}_{i,j} - \Theta_{i,j})^2] = \mathrm{Var}(\hat{\Theta}_{i,j})
   \]

   The expected value of the sum of squared errors is:
   \[
   \mathbb{E}\left[\sum_{i=1}^{n} \sum_{j=1}^{n+m} (\hat{\Theta}_{i,j} - \Theta_{i,j})^2\right] = \sum_{i=1}^{n} \sum_{j=1}^{n+m} \mathrm{Var}(\hat{\Theta}_{i,j}) = \mathbf{tr}(\mathrm{Cov}(\hat{\Theta}))
   \]

The mean squared error (MSE) is:
   \[
   \mathrm{MSE} = \frac{1}{n(n+m)} \sum_{i=1}^{n} \sum_{j=1}^{n+m} (\hat{\Theta}_{i,j} - \Theta_{i,j})^2
   \]
   Taking the expectation:
   \[
   \mathbb{E}[\mathrm{MSE}] = \frac{1}{n(n+m)} \mathbb{E}\left[\sum_{i=1}^{n} \sum_{j=1}^{n+m} (\hat{\Theta}_{i,j} - \Theta_{i,j})^2\right] = \frac{1}{n(n+m)} \mathbf{tr}(\mathrm{Cov}(\hat{\Theta}))
   \]

The RMSE is the square root of the MSE which is then:
   \[
   \mathrm{RMSE} = \sqrt{\mathrm{MSE}} = \sqrt{\frac{1}{n(n+m)} \mathbf{tr}(\mathrm{Cov}(\hat{\Theta}))}
   \]

\subsection{Objective Equivalence}
Here we prove that minimizing the trace of $\mathrm{Cov}(\hat{\Theta})$ is equivalent to minimizing the trace of \(\Gamma = \sigma^2 \left(ZZ^{\top}\right)^{-1}\). Given that $$\mathrm{Cov}(\hat{\Theta}) = \mathrm{I}_n \otimes \Gamma.$$

Given two square matrices \(A\) and \(B\), the trace of the Kronecker product is the product of their traces:
   \[
   \mathbf{tr}(A \otimes B) = \mathbf{tr}(A) \cdot \mathbf{tr}(B).
   \]

Applying this property to \(\mathrm{Cov}(\hat{\Theta}) = \mathrm{I}_n \otimes \Gamma\), we have:
   \[
   \mathbf{tr}(\mathrm{Cov}(\hat{\Theta})) = \mathbf{tr}(\mathrm{I}_n \otimes \Gamma).
   \]

Since \(\mathrm{I}_n\) is the \(n \times n\) identity matrix, its trace is \(n\):
   \[
   \mathbf{tr}(\mathrm{I}_n) = n.
   \]

   Therefore,
   \[
   \mathbf{tr}(\mathrm{I}_n \otimes \Gamma) = \mathbf{tr}(\mathrm{I}_n) \cdot \mathbf{tr}(\Gamma) = n \cdot \mathbf{tr}(\Gamma).
   \]

   The total variance of the estimates of \(\hat{A}\) and \(\hat{B}\) as measured by \(\mathbf{tr}(\mathrm{Cov}(\hat{\Theta}))\) is:
   \[
   \mathbf{tr}(\mathrm{Cov}(\hat{\Theta})) = n \cdot \mathbf{tr}(\Gamma).
   \]

   Since \(n\) is a constant factor (the dimension of the state space), minimizing \(\mathbf{tr}(\mathrm{Cov}(\hat{\Theta}))\) is equivalent to minimizing \(\mathbf{tr}(\Gamma)\). This is an important result as it reduces the optimization matrix size from $n(n+m) \times n(n+m)$ to $(n+m) \times (n+m)$.

\subsection{Linear Program Heuristic}
We proposed that instead of minimizing $\mathbf{tr} (\hat{W}^{-1})$ we could instead minimize $- \mathbf{tr}(\hat{W})$ which converts the problem from a semidefinite program to a linear program allowing for even more efficient solution time. As previously mentioned, minimizing \(\mathbf{tr}(W^{-1})\) penalizes small eigenvalues of \(W\), ensuring that estimation errors are balanced across all directions. In contrast, the \(-\mathbf{tr}(W)\) objective can result in a few large eigenvalues dominating the objective, while others remain small. As a result, this could concentrate information gain in a single direction, resulting in poor performance in orthogonal directions. 

Consider the hypothetical example where there is no state involved so $\Gamma = \hat \sigma^2 (\Upsilon \Upsilon^{\top})^{-1}$. This is ordinary experiment design. Using the linear program objective, we should maximize $\mathbf{tr} (\hat \sigma^{-2} (\Upsilon \Upsilon^{\top}))$.  The solution to this problem would use a constant input $u$, the one that maximizes $\|u\|_2$ subject to $u \in \mathcal U$ (the set of allowable controls). Unless you have prior information, that’s the worst possible choice of input. It gives excellent performance in one direction, and the worst possible in all others (i.e., orthogonal to that direction). This highlights the main theoretical drawback of the linear program method, while the semidefinite program formulation is more widely applicable to different problem instances.  

\subsection{Additional Results}
\begin{figure}
\centering
\includegraphics[width=\textwidth]{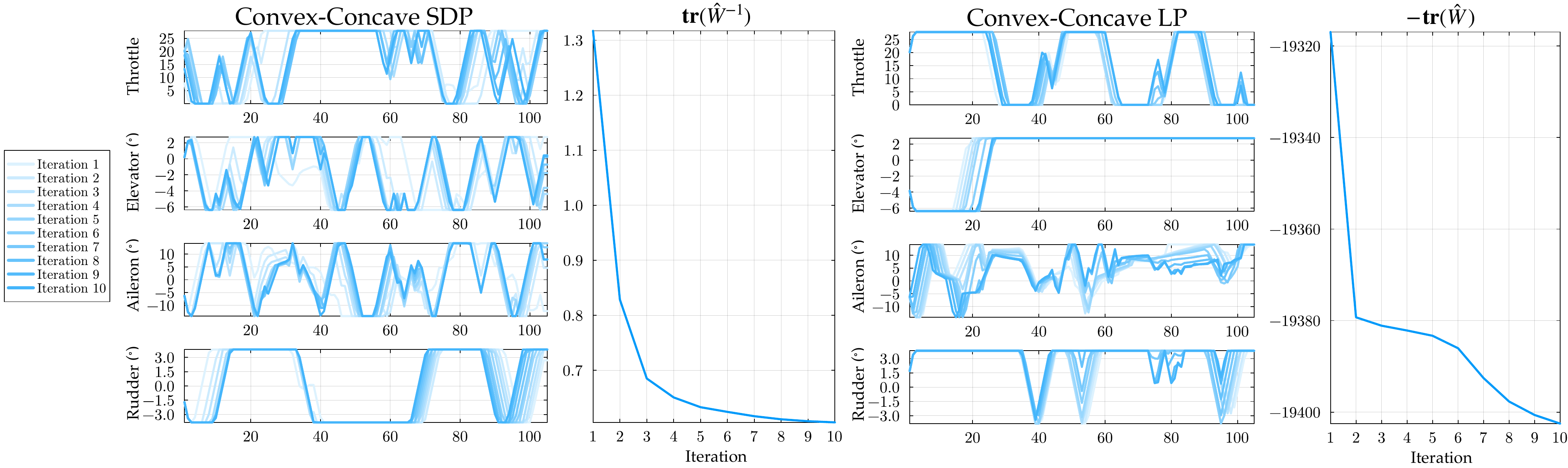}
\caption{Optimization progression of the control inputs using both the SDP and LP formulation with the corresponding objective values shown for the F-16 example using the Aerobench simulator.}
\end{figure}

\begin{table}[h!]
\centering
\caption{Average Relative Peak Factor and Maximum Differences (Slew Rate)}
\begin{tabular}{lccccccccc}
 & \multicolumn{4}{c}{RPF} && \multicolumn{4}{c}{Maximum Difference} \\
 \cline{2-5}
 \cline{7-10}
Control Input & SDP & LP & Orthogonal & Random && SDP & LP & Orthogonal & Random \\
\hline 
Throttle & 0.759 & 0.759 & 0.759 & 0.759 && 4.241 & 4.438 & 4.443 & 4.526 \\
Elevator & 0.786 & 0.779 & 0.765 & 0.759 && 0.246 & 0.222 & 0.219 & 0.224 \\
Aileron & 0.843 & 0.829 & 0.782 & 0.765 && 0.300 & 0.222& 0.238 & 0.226  \\
Rudder & 0.842 & 0.836 & 0.792 & 0.765 && 0.244 & 0.222 & 0.23 & 0.226  \\
\end{tabular}
\label{table:merged}
\end{table}

\end{document}